%% file: order.tex
\documentclass[draft, 12pt, journal, onecolumn]{IEEEtran}
\usepackage[utf8]{inputenc}
\usepackage{amsmath}
\usepackage{amsfonts}
\usepackage{amssymb}
\usepackage{amsthm}
\newtheorem{theorem}{Theorem}
\newtheorem{lemma}{Lemma}
\newtheorem{corollary}{Corollary}

\newtheorem{definition}{Definition}
\newtheorem{proposition}{Proposition}
\newtheorem{example}{Example}
\usepackage[final]{graphicx}
\usepackage{caption}
\usepackage{subcaption}
\newcommand{\expectation}{\ensuremath{\mathbb{E}}}
\newcommand{\Expt}{\expectation}
\newcommand{\probability}{\ensuremath{\mathbb{P}}}
\newcommand{\Prob}{\probability}

\def\fF{\mathbb{F}}

\def\cY{\mathcal{Y}}

\def\N{\mathbb{N}}
\def\cY{\mathcal{Y}}

\def\<{\langle}
\def\>{\rangle}

\begin{document}

\title{The Symmetric Convex Ordering: \\ A Novel Partial Order for B-DMCs ordering the Information Sets of Polar Codes}
\author{\IEEEauthorblockN{Mine Alsan}\\
\small\IEEEauthorblockA{Email: minealsan@gmail.com}
\normalsize
\thanks{The material in this paper was presented in part at the IEEE International Symposium on Information Theory, Honolulu, USA, July 2014.}}
\maketitle
\IEEEpeerreviewmaketitle

\maketitle

\begin{abstract}
In this paper, we propose a novel partial order for binary discrete memoryless channels that we call the symmetric convex ordering. We show that Ar{\i}kan's polar transform preserves `symmetric convex orders'. Furthermore, we show that while for symmetric channels this ordering turns out to be equivalent to the stochastic degradation ordering already known to order the information sets of polar codes, a strictly weaker partial order is obtained when at least one of the channels is asymmetric. In between, we also discuss two tools which can be useful for verifying this ordering: a criterion known as the cut criterion and channel symmetrization. Finally, we discuss potential applications of the results to polar coding over non-stationary channels.
\end{abstract}

\begin{IEEEkeywords}
Polar coding, partial orders, convex ordering, increasing convex ordering, non-stationary memoryless channels. 
\end{IEEEkeywords}

\section{Introduction}
To set up, let us think about the elegant principle behind the construction of the information set of a polar code for a given binary discrete memoryless channel (B-DMC) $W:\fF_2\to\cY$. The construction starts by combining and splitting independent copies of the channel by applying Ar{\i}kan's polar transform \cite[Eqs (17) and (18)]{1669570} in a recursive fashion. In the first stage, two new binary input channels $W^-:\fF_2\to\cY^2$ and $W^+:\fF_2\to\cY^2\times\fF_2$ are synthesized from two independent copies of $W$. Then, applying the polar transform to these new channels, the channels $W^{--}:=(W^-)^-$, $W^{-+}:=(W^-)^+$, $W^{+-}:=(W^+)^-$, and $W^{++}:=(W^+)^+$ are obtained. More generally, the repeated application yields, at stage $n = 1, 2, \ldots$, a set of $2^n$ channels
\begin{equation}\label{eq:ch_pol_proc}
\bigl\{W^{s^n}:s^n\in\{+,-\}^n\bigr\}.
\end{equation}
After a long sequence of such operations, the synthesized channels cluster eventually in two states: almost perfect or completely noisy \cite[Theorem 1]{1669570}. As the main idea behind the construction of the information set is to ensure that the overall error probability 
of the decoding procedure is small, the information set of a polar code of block-length $N = 2^n$ for the channel $W$, denoted as $\mathcal{A}_{N}(W)$, is specified by picking
from the set $\{+,-\}^n$ the indices of the synthetic channels which are good for uncoded transmission, i.e., the code designer is looking for the set
\begin{equation}\label{eq:vague_def_info_set}
\mathcal{A}_{N}(W) = \bigl\{s^n\in\{+,-\}^n: W^{s^n} \hbox{ is `\textit{good}'} \bigr\}.
\end{equation}

Although the information set is explicitly defined, a difficulty arises in accomplishing the computation of the exact transition probabilities of the synthetic channels  as their output alphabets grow larger and larger with the application of the polar transform. Initially, this problem was solved in \cite{1669570} by proposing to approximate the computations by estimating the good channels via their Bhattacharyya distance with the help of the Monte Carlo method. 
Though the computations could be carried offline using the channel model at hand, this approach
had two limitations: complexity and reliability of the Monte Carlo estimates. The problem of finding an efficient code construction algorithm for polar codes
was first addressed by Mori and Tanaka in \cite{5205857} and \cite{5166430}. Thanks to Tal and Vardy, an algorithm to carry the computations approximately (but within guaranteed bounds) and efficiently was thought out later in \cite{6557004}. 
Other methods such as the Gaussian approximation for computing the bit error probabilities over Gaussian channels \cite{6279525} have been also proposed. 
Overall, the fact that polar codes can be explicitly defined and also efficiently constructed are important for the practice of polar coding. 

Another `not so hidden' characteristic of \eqref{eq:vague_def_info_set} is the reliance of the definition on a specific channel. This apparent observation led to a question of both theoretical and practical interest after the invention of polar codes: How large is $\mathcal{A}_{N}(W)\cap \mathcal{A}_{N}(V)$, 
for two given channels $W$ and $V$. Two partial orders have been pointed out in~\cite{1669570} which order the information sets of polar codes: 
Any binary erasure channel provides good indices for all other B-DMCs having smaller Bhattacharyya parameters, 
and any channel which is degraded with respect to another B-DMC provides good indices for the upgraded channel\footnote{$W$ is upgraded with respect to $V$ if and only if $V$ is degraded with respect to $W$.}. 
In this paper, we will show that these partial orderings can be studied in the context of a stochastic order known as convex ordering. Interestingly, it will turn out that the solution to the efficient computation problem found in \cite{6557004} is closely tied to the notion of convex ordering. 

\section{Overview of the Results}
Many channel parameters can be used to quantify ``good'' in \eqref{eq:vague_def_info_set}.
Originally, Ar{\i}kan chose to define the information sets of polar codes in terms of the Bhattacharyya parameters of the synthetic channels. Letting 
\begin{equation}\label{eq:Bhattacharyya}
Z(W) = \displaystyle\sum_{y}\sqrt{W(y|0)W(y|1)}
\end{equation}
denote the Bhattacharyya parameter of a B-DMC $W$, \cite{1669570} gives the following definition:
\begin{equation}
\mathcal{A}_{N}^\epsilon(W) = \bigl\{s^n\in\{+,-\}^n: Z(W^{s^n}) \leq \epsilon \bigr\},
\end{equation}
for $\epsilon\in(0, 1)$. $Z(W)$ is a well know upper bound to the average maximum likelihood decoding error probability $P_{\textnormal{e, ML}}(W)$ of a single bit transmission over the channel $W$. Thus, when the Bhattacharyya parameter of a synthetic channel is small, the resulting error probability $P_{\textnormal{e, ML}}(W^{s^n})$ is also small and the channel is good for uncoded transmission. Following this reasoning, it is not difficult to see that any channel parameter appearing in a meaningful upper bound to $P_{\textnormal{e, ML}}(W)$ is eligible, as this bound would apply individually to the synthetic channels and would serve to upper bound the successive cancellation decoding error probability of polar codes via the union bound\footnote{See \cite[Section V]{1669570} for an analysis of the error performance of polar codes.}. For instance, the symmetric capacity of a channel, defined as 
\begin{equation}\label{eq:sym_cap}
I(W) = \displaystyle\sum_{x, y} \frac{1}{2} W(y\mid x) \log{\frac{W(y\mid x)}{\frac{1}{2} W(y\mid 0) + \frac{1}{2} W(y\mid 1)}},
\end{equation}
is another possible choice used to quantify the information sets of polar codes, see  \cite{6033892}.

In this paper, we will consider a family of such quantifiers generated by the following class of functions:
\footnotetext[3]{A function $f(\delta)$ is called symmetric if $f(\delta) = f(-\delta)$, for all $\delta\in\mathbf{R}$.}
\begin{multline}\label{eq:family_cx_s}
 \mathcal{F}_{cx, s} = \{f_s:[-1, 1]\to [0, 1]; \hspace{1mm} f_s 
 \hbox{ is symmetric\footnotemark[3] and convex } \\
 \hbox{such that }f_s(0)=0 \hbox{ and }f_s(1)=1\}.
\end{multline}
The functions in $\mathcal{F}_{cx, s}$ will take as argument the following channel parameter: 
\begin{equation}\label{eq:delta}
\Delta_W(y) = \frac{W(y\mid0)-W(y\mid1)}{W(y\mid0)+W(y\mid1)}.
\end{equation} 

Let us start by demonstrating the generated upper bounds. We first consider the absolute value function from the class. Suppose the inputs of $W$ are used with equal frequency. Then, the output distribution of the channel is given by
\begin{equation}\label{eq:dist}
q_W(y)=\frac{W(y\mid0) + W(y\mid1)}{2},
\end{equation}
for $y\in\mathcal{Y}$, and the expectation of $\lvert\Delta_W(Y)\rvert$ evaluated under $q_{W}(y)$ gives
\begin{equation}
E[\lvert\Delta_W(Y)\rvert] = \displaystyle\frac{1}{2}\displaystyle\sum_{y}\lvert W(y|0) - W(y|1)\rvert.
\end{equation}
Thus, this expectation computes the variational distance between the channel's transition probabilities $W(y|0)$ and $W(y|1)$. We denote this variational distance by $T(W)\triangleq E[\lvert\Delta_W(Y)\rvert]$, and note that it is related to the error probability as follows \cite[e.g. Eq. 47]{1089532}: 
  \begin{equation}
 T(W) 
% \nonumber\\
%&= \displaystyle\frac{1}{2} \left( \Prob_{W(.|0)}\left[L_W \lneq 1 \right] - \Prob_{W(.|0)}\left[L_W \gneq 1 \right] \right.\nonumber\\ 
%&\hspace{10mm}- \left.\Prob_{W(.|1)}\left[L_W \lneq 1 \right] + \Prob_{W(.|1)}\left[L_W \gneq 1 \right] \right) \nonumber\\
%&= \displaystyle\frac{1}{2} \left(1 - 2\Prob_{W(.|0)}\left[L_W \gneq 1 \right] + 1 - 2\Prob_{W(.|1)}\left[L_W \lneq 1 \right] \right) \nonumber\\
\label{eq:var_dist_ML_err}= 1 - 2 P_{\textnormal{e, ML}}(W).
\end{equation}
So, to have a small $P_{\textnormal{e, ML}}(W)$, the channel at hand must
have a large variational distance (close to 1). Equivalently,
it would be sufficient that 
\begin{equation*}
T_{f_s}(W) \triangleq \Expt\left[f_s\left(\Delta_W\right)\right] 
\end{equation*}
is large for any $f_s\in\mathcal{F}_{cx, s}$, since
\begin{equation*}
T(W) \geq T_{f_s}(W)
\end{equation*}
always holds. Upon noticing that $T_{f_s}(W)\in[0, 1]$, we conclude via \eqref{eq:var_dist_ML_err} that the parameters $T_{f_s}(W)$ generate a family of upper bounds to $P_{\textnormal{e, ML}}(W)$. 

Based on this observation, we generalize the definition of the information sets of polar codes as follows. 
\begin{definition}\label{def:info_set}
Let $f_s\in\mathcal{F}_{cx, s}$ and $\epsilon\in(0, 1)$. 
$W$ is called `$\epsilon$-good' if $T_{f_s}(W) \geq 1-\epsilon$ holds. Accordingly, the information set definition is adapted as
\begin{equation}\label{eq:gen_info_set}
\mathcal{A}_{N}^{f_s, \epsilon}(W) = \bigl\{s^n\in\{+,-\}^n: T_{f_s}(W^{s^n}) \geq 1-\epsilon \bigr\}.
\end{equation}
\end{definition}
For instance, the particular choice of $f_s(\delta) = 1 - h(\frac{1+\delta}{2})$, where $h(.)$ denotes the binary entropy function, or $f_s(\delta) = 1 - \sqrt{1-\delta^2}$
lead to information set definitions based on the values of the symmetric capacities and 
the Bhattacharyya parameters of the synthetic channels, respectively.

In Section \ref{sec:nov_order}, we will show that, in essence, taking $\Delta_W(Y)$ as argument,  the class of symmetric convex functions generates a partial ordering for B-DMCs which orders the information sets of polar codes: 
\begin{equation*}
\mathcal{A}_{N}^{f_s, \epsilon}(V) \subseteq \mathcal{A}_{N}^{f_s, \epsilon}(W), \forall N, \forall \epsilon \quad \hbox{if} \quad T_{f_s}(V) \leq T_{f_s}(W),  \forall f_s\in\mathcal{F}_{cx, s}.
\end{equation*}
This result will follow as a corollary to Theorem \ref{thm:icx} which will show that the polar transform preserves symmetric convex orderings.

Note that Theorem \ref{thm:icx} will be stated for a slightly more general polar transform, denoted by $\<W_1,W_2\>^\pm$, that synthesizes two channels from two independent (but not necessarily identical) binary input channels $W_1:\fF_2\to\cY_1$ and $W_2:\fF_2\to\cY_2$. Given two such channels, the generalized polar transform synthesizes the channels $W_{1,2}^- = \<W_1,W_2\>^-:\fF_2\to\cY_1\times\cY_2$ and $W_{1,2}^+ = \<W_1,W_2\>^+:\fF_2\to\cY_1\times\cY_2\times\fF_2$ with transition probabilities given by
\begin{align}
\label{align:trans1} & W_{1,2}^-(y_1y_2|u_1) =
	\sum_{u_2\in\fF_2}\tfrac12 W_1(y_1|u_1\oplus u_2) W_2(y_2|u_2), \\
\label{align:trans2} &W_{1,2}^+(y_1 y_2 u_1|u_2) =
	\tfrac12 W_1(y_1 | u_1 \oplus u_2) W_2(y_2 | u_2). 
\end{align}

Once the theorem will be proved, we will compare in Section \ref{sec:exploration} the symmetric convex ordering with the stochastic degradation ordering already known to order the information sets
of polar codes: We will show that while for symmetric channels this ordering is equivalent to stochastic degradation, 
a strictly weaker partial order is obtained when at least one of the channels is asymmetric. 
We will illustrate this by a particular example which studies both orderings between a Z-channel and a binary symmetric channel whose inputs are used with equal frequency. 
In the process, we will also present tools which can be useful for verifying the symmetric convex ordering: the cut criterion due to \cite{stanford1962generalized} and channel symmetrization. 

In the following Section \ref{sec:pol_property}, Lemma \ref{lem:pol_property} will demonstrate that after the generalized polar transform is applied, the created channels and the original channels are ordered with the symmetric convex ordering. Next, Section \ref{sec:applications} will present two additional applications of the new ordering to polar coding over non-stationary B-DMCs which was recently studied in \cite{6874843}. We will first discuss how the symmetric convex ordering can be useful for efficiently constructing polar codes in this scenario. Then, we will show that Theorem \ref{thm:icx} is also helpful when dealing with the problem of universal polar coding with channel knowledge at the decoder over non-stationary channels. Finally, Section \ref{sec:final-rem} will interpret the results in view of the state-of-the-art literature.

\section{A Novel Partial Ordering for B-DMCs}\label{sec:nov_order}
First and foremost, we designate the novel ordering. We use capital letters to denote random variables and lower-case letters to denote their realizations.
\begin{definition}\label{def:sym_cx_ordering}
We say that two B-DMCs $W$ and $V$ satisfy the \textit{symmetric convex ordering} if
\begin{equation*}
\Expt\left[f_s(\Delta_V)\right] \leq \Expt\left[f_s(\Delta_W)\right],
\end{equation*}
for all functions $f_s\in\mathcal{F}_{cx, s}$. 
\end{definition}

Next, we bridge this definition with a well known stochastic order. Let $\Delta_1$ and  $\Delta_2$ be two random variables with distributions $F_{\Delta_1}$ and $F_{\Delta_2}$, respectively.
\begin{definition}\label{def:icx}\cite{szekli1995stochastic}
$\Delta_1$ is smaller with respect to the \textit{increasing convex ordering} (\textit{decreasing concave ordering}) than $\Delta_2$, written $\Delta_1 \prec_{icx} \Delta_2$ ($\Delta_1 \prec_{dcv} \Delta_2$), if 
\begin{equation}\label{eq:icx_ineq_cond}
\Expt\left[f(\Delta_1)\right] \leq \Expt\left[f(\Delta_2)\right],
\end{equation}
for all increasing convex (decreasing concave) functions $f$ for which the expectations exist.
\end{definition}
As any result involving the $\prec_{icx}$ ordering can be mapped to the $\prec_{dcv}$ ordering, we will stick to the first one. 
Alternatively, the $\prec_{icx}$ ordering can be described by using only the class of symmetric functions.
\begin{proposition}\label{prop:sym_order_equiv}
$\lvert \Delta_1 \rvert \prec_{icx} \lvert \Delta_2\rvert$ if and only if
\begin{equation*}
\Expt\left[f_s(\Delta_1)\right] \leq \Expt\left[f_s(\Delta_2)\right],
\end{equation*}
for all convex symmetric functions $f_s$ for which the expectations exist.
\end{proposition}
\begin{proof}
The proof follows by the fact that $f_s(\lvert \delta \rvert) = f_s(\delta)$ holds for any symmetric function $f_s(\delta)$, $\delta\in\mathbf{R}$.
\end{proof}
Thus, the new partial ordering introduced in Definition \ref{def:sym_cx_ordering} 
is an increasing convex ordering for the absolute value of the channels' $\Delta_W$ parameters. 

Now, we are ready to state the main result.
\begin{theorem}\label{thm:icx} 
Let $W_1$, $W_2$, $V_1$, and $V_2$ be B-DMCs such that 
\begin{equation*}
\lvert\Delta_{V_1}\rvert \prec_{icx} \lvert\Delta_{W_1}\rvert
\quad \hbox{and} \quad \lvert\Delta_{V_2}\rvert \prec_{icx} \lvert\Delta_{W_2}\rvert
\end{equation*} 
hold. Then, the polar transform preserves this ordering, i.e., $\big\lvert\Delta_{V_{1,2}^\pm}\big\rvert \prec_{icx} \big\lvert\Delta_{W_{1,2}^\pm}\big\rvert$.
\end{theorem}
\begin{proof}[Proof of Theorem \ref{thm:icx}]
We will use the characterization given in Proposition \ref{prop:sym_order_equiv} in the proof. 
After applying the polar transform to the channels, one can derive the following recursion
\begin{align}
\label{align:delta_rec_minus} &\Delta_{W_{1,2}^-}(Y_1Y_2) = \Delta_{W_1}(Y_1)\Delta_{W_2}(Y_2), \\
\label{align:delta_rec_plus} &\Delta_{W_{1,2}^+}(Y_1Y_2U_1) = \displaystyle\frac{\Delta_{W_1}(Y_1)+ (-1)^{U_1} \Delta_{W_2}(Y_2)}{1+ (-1)^{U_1} \Delta_{W_1}(Y_1)\Delta_{W_2}(Y_2)},
\end{align}
where $Y_1Y_2\sim q_{W_1}(y_1)q_{W_2}(y_2)$, and
$$Y_1Y_2U_1 \sim q_{W_1}(y_1)q_{W_2}(y_2)\displaystyle\frac{1+ (-1)^{u_1} \Delta_{W_1}(y_1)\Delta_{W_2}(y_2)}{2}.$$
See \cite[Proofs of Lemmas 1 and 2]{6731577} for a proof.

Let $f_s(\delta)$ be a function which is convex and symmetric in $\delta\in[-1, 1]$. Note that by the convexity and the symmetry assumptions, the function will be increasing in $\delta\in[0, 1]$. 
For the minus polar transform, we write
\begin{align*}
 &\displaystyle\sum_{y_1y_2} q_{W_{1,2}^-}(y_1y_2) f_s\left(\Delta_{W_{1,2}^-}(y_1y_2)\right) \\
% = &\displaystyle\sum_{y_1} q_{W_1}(y_1) \displaystyle\sum_{y_2} q_{W_2}(y_2)
% f\left(\Delta_{W_1}(y_1) \Delta_{W_2}(y_2)\right) \\
 &\hspace{20mm}= \displaystyle\sum_{y_1} q_{W_1}(y_1) \displaystyle\sum_{y_2} q_{W_2}(y_2)
 f^-\left(\Delta_{W_1}(y_1), \Delta_{W_2}(y_2)\right) 
\end{align*}
where ${f^-}\left(\delta_1, \delta_2\right) = f_s\left(\delta_1 \delta_2\right)$, for $\delta_1, \delta_2\in[-1, 1]$. As we assumed $f_s(\delta)$ to be convex and symmetric in its argument, so is $f^-$ in both of its arguments. 
Similarly for the plus polar transform, we write
\begin{align*}
 &\displaystyle\sum_{y_1y_2u_1} q_{W_{1,2}^+}(y_1y_2u_1) f_s\left(\Delta_{W_{1,2}^+}(y_1y_2u_1)\right) \\
 &\hspace{20mm}= \displaystyle\sum_{y_1y_2} q_{W_{1,2}^+}(y_1y_20) f_s\left(\Delta_{W_{1,2}^+}(y_1y_2 0)\right) + \displaystyle\sum_{y_1y_2} q_{W_{1,2}^+}(y_1y_2 1) f_s\left(\Delta_{W_{1,2}^+}(y_1y_2 1)\right) \\
 &\hspace{20mm}= \displaystyle\sum_{y_1} q_{W_1}(y_1) \displaystyle\sum_{y_2} q_{W_2}(y_2)  f^+\left(\Delta_{W_1}(y_1), \Delta_{W_2}(y_2)\right), 
\end{align*}
where
\begin{equation}\label{eq:f_plus}
f^+\left(\delta_1, \delta_2\right) 
= \displaystyle\frac{1 + \delta_1\delta_2}{2} f_s\left(\displaystyle\frac{\delta_1 + \delta_2}{1 + \delta_1\delta_2}\right) 
+ \displaystyle\frac{1 - \delta_1\delta_2}{2} f_s\left(\displaystyle\frac{\delta_1 - \delta_2}{1 - \delta_1\delta_2}\right),
\end{equation}
for $\delta_1, \delta_2\in[-1, 1]$.
Lemma \ref{lem:f_plus_convex} in the Appendix shows that $f^+$ is also a convex and symmetric function in both of its arguments.  

So, using the assumptions $\lvert \Delta_{V_1} \rvert \prec_{icx} \lvert \Delta_{W_1} \rvert$ and $\lvert \Delta_{V_2} \rvert \prec_{icx} \lvert \Delta_{W_2} \rvert$, we deduce that  
\begin{align*}
 &\displaystyle\sum_{y_1} q_{V_1}(y_1)\displaystyle\sum_{y_2}q_{V_2}(y_2) f^{\pm}\left(\Delta_{V_1}(y_1), \Delta_{V_2}(y_2)\right) \\
&\hspace{10mm}\leq \displaystyle\sum_{y_1} q_{V_1}(y_1)\displaystyle\sum_{y_2}q_{W_2}(y_2) f^{\pm}\left(\Delta_{V_1}(y_1), \Delta_{W_2}(y_2)\right) \\
&\hspace{10mm}= \displaystyle\sum_{y_2} q_{W_2}(y_2)\displaystyle\sum_{y_1}q_{V_1}(y_1) f^{\pm}\left(\Delta_{V_1}(y_1), \Delta_{W_2}(y_2)\right) \\
&\hspace{10mm}\leq \displaystyle\sum_{y_2} q_{W_2}(y_2)\displaystyle\sum_{y_1}q_{W_1}(y_1) f^{\pm}\left(\Delta_{W_1}(y_1), \Delta_{W_2}(y_2)\right).
\end{align*}
This proves our claim that both $\big\lvert \Delta_{V_{1,2}^\pm} \big\rvert \prec_{icx} \big\lvert \Delta_{W_{1,2}^\pm} \big\rvert$ hold.
\end{proof}
Using the generalized information set definition given in \eqref{eq:gen_info_set}, we get the following corollary to the previous theorem.
\begin{corollary}\label{cor:info_set}
Let $W$ and $V$ be two B-DMCs which satisfy the symmetric convex ordering as defined in \ref{def:sym_cx_ordering}. Then, 
\begin{equation}\label{eq:cor_order_info}
\mathcal{A}_{N}^{f_s, \epsilon}(V) \subseteq \mathcal{A}_{N}^{f_s, \epsilon}(W), 
\end{equation}
holds for all $f_s\in\mathcal{F}_{s, cx}$ and for all $N = 2^n$ with $n=1, 2, \ldots$.
\end{corollary}
\begin{proof}
The assumption on the channels implies via Proposition \ref{prop:sym_order_equiv} that 
$\lvert\Delta_{V}\rvert \prec_{icx} \lvert\Delta_{W}\rvert$ holds. Then, \eqref{eq:cor_order_info} follows by Theorem \ref{thm:icx}.  
\end{proof}

As we pointed out earlier, it is stated in \cite{1669570} that the information sets of polar codes are ordered for stochastically degraded channels. 
See \cite[Lemma 4.7]{4461/THESES} for a proof of the fact that stochastic degradation is preserved under the original polar transform and \cite[Appendix 2.B]{ALSAN/THESES} for the fact that two stochastically degraded DMCs are ordered in their $E_0(\rho)$ parameters for any $\rho>0$. (Note that for any fixed $\rho>0$, $E_0(\rho)$ can be expressed as the minus logarithm of the expectation of a function belonging to $\mathcal{F}_{s, cx}$ \cite[Eq. (8) and Lemma 4]{6731577}). It would therefore be of interest to compare the symmetric convex ordering we introduced with stochastic degradation. 

\section{Exploration}\label{sec:exploration}
\subsection{Convex Ordering}\label{subsec:convex}
The material up to and including Theorem \ref{thm::cut_criterion} is drawn from \cite[Section 1.3]{szekli1995stochastic}. 
The following definition introduces a special case of the increasing convex ordering. 
\begin{definition}\cite[Theorem B]{szekli1995stochastic}\label{def:cx}
Suppose $\Delta_1$ and $\Delta_2$ have equal mean values. 
$\Delta_1$ is smaller with respect to the convex ordering than $\Delta_2$, written $\Delta_1 \prec_{cx} {\Delta_2}$, if and only if 
\begin{equation*}
\Expt\left[f(\Delta_1)\right] \leq \Expt\left[f(\Delta_2)\right],
\end{equation*}
for all convex $f$ for which the expectations exist.
\end{definition} 

\begin{definition}\cite{szekli1995stochastic}
A Markov kernel is a function $T_M(\delta, E)$, $\delta\in\mathbf{R}$, $E\in\mathbf{B}$, such that  $T_M(\delta, .)$ is a 
probability measure on $\mathbf{R}$ for each fixed $\delta$ and $T_M(., E)$ is a measurable function for each fixed $E$. $T_M$ is 
mean value preserving if the mean value of the probability measure $T_M(\delta, .)$ is equal to $\delta$. 
%$\Expt[Y|X=d] = d$ for $\Prob(Y\in E|X=d) = T_M(d, E)$.
\end{definition}

An alternative description of convex ordering due to Blackwell \cite{Blackwell1953} is given in \cite[Theorem C]{szekli1995stochastic}. Below is the statement of this theorem.
%We will see later that this result establishes the equivalence between stochastic degradation and convex ordering.
\begin{theorem}\cite{Blackwell1953}\label{thm:Markov_kernel}
$\Delta_1\prec_{cx}\Delta_2$ if and only if there exists a mean value preserving Markov kernel $T_M$ such that $F_{\Delta_2} = T_MF_{\Delta_1}$, i.e.,
\begin{equation*}
F_{\Delta_2}(\delta_2) = \Expt\big[T_M\big(\Delta_1, \left(-\infty, \delta_2\right]\big)\big].
\end{equation*}
\end{theorem}

\begin{definition}
A random variable $\Delta$ is called symmetric if the distribution of $\Delta$ satisfies
$F_\Delta(\delta) = 1 - F_\Delta(-\delta)$, for all $\delta\in \mathbf{R}$.
\end{definition}
In the next proposition, we exploit this symmetry property.

\begin{proposition}\label{prop:cx_order_sym}
For symmetric $\Delta_1$ and $\Delta_2$, $\Delta_1 \prec_{cx} \Delta_2$ 
if and only if $\lvert \Delta_1 \rvert \prec_{icx} \lvert \Delta_2\rvert$.
\end{proposition}
\begin{proof}
The `only if part' follows by definition. So, we only need to prove the `if part'. Let $f(\delta)$ be a convex function in $\delta\in \mathbf{R}$. As $\Delta_1$ is symmetric, we can write
\begin{equation*}
\Expt\left[f(\Delta_1)\right] = \Expt\left[\displaystyle\frac{f(\Delta_1) + f(-\Delta_1)}{2} \right]
= \Expt\left[f_s(\Delta_1)\right],
\end{equation*}
where $f_s(\delta) = \left(f(\delta) + f(-\delta)\right)/2$ is a convex symmetric function. In particular, $f_s(.)$ is increasing on $\mathbf{R_{+}}$. Hence 
using $\lvert \Delta_1 \rvert \prec_{icx} \lvert \Delta_2\rvert$, we get
\begin{equation*}
\Expt\left[f(\Delta_1)\right] = \Expt\left[f_s(\lvert \Delta_1 \rvert)\right] \leq \Expt\left[f_s(\lvert \Delta_2 \rvert)\right] = \Expt\left[f(\Delta_2)\right]. \qedhere
\end{equation*}
\end{proof}

Now, we show that for symmetric channels convex ordering is equivalent to stochastic degradation. 
Let $V:\fF_2\to\cY$ be stochastically degraded with respect to $W:\fF_2\to\cY$. Then, by definition, there exists a channel $P: \mathcal{Y}\to\mathcal{Y}$ such that 
\begin{equation}\label{eq:sd}
 V(y|x) = \displaystyle\sum_{z\in\mathcal{Y}} W(z|x) P(y|z)
\end{equation}
hold for all $y\in\mathcal{Y}$. In this case, one can derive the following:
\begin{equation*}
 \Delta_V(y) = \displaystyle\frac{V(y|0) - V(y|1)}{V(y|0) + V(y|1)} = \displaystyle\sum_{z} \bar{P}(z|y) \Delta_W(z),
\end{equation*}
where
\begin{equation*}
 \bar{P}(z|y) = \displaystyle\frac{q_W(z)P(y|z)}{\displaystyle\sum_z q_W(z)P(y|z)}
\end{equation*}
corresponds to the inputs posterior probabilities given the output of the channel $P$. 
So, for any convex function $f(.)$, we obtain
\begin{align}
\Expt\left[f(\Delta_V)\right] = &\displaystyle\sum_{y} q_V(y) f(\Delta_V(y)) \nonumber\\
 = &\displaystyle\sum_{y} \left(\displaystyle\sum_{z}q_W(z)P(y|z)\right)  f\left(\displaystyle\sum_{z} \bar{P}(z|y) \Delta_W(z)\right) \nonumber\\
 \leq &\displaystyle\sum_{y} \displaystyle\sum_{z}q_W(z)P(y|z)f(\Delta_W(z)) \nonumber\\
 \label{eq::deg_convex}= &\displaystyle\sum_{z}q_W(z)f(\Delta_W(z)) = \Expt\left[f(\Delta_W)\right], 
\end{align}
where the inequality follows by Jensen's inequality. 
In particular, the ordering holds with equality for the function $f(\delta) = \delta$. 
Hence, degradation preserves the mean value, i.e., $E[\Delta_{W}] = E[\Delta_{V}]$. 
By Definition \ref{def:cx}, we conclude the order relation $\Delta_{V} \prec_{cx} \Delta_{W}$ holds for stochastically degraded channels. 

To show the reverse implication, suppose the channels satisfy $\Delta_{V} \prec_{cx} \Delta_{W}$. By Theorem \ref{thm:Markov_kernel}, 
there exists a Markov kernel $T_M$ such that
\begin{align}
\label{align:MK_1}&\displaystyle\sum_{z} T_M(y, z) = 1,  \\
\label{align:MK_2}&\Delta_V(y) = \displaystyle\sum_{z} T_M(y, z) \Delta_W(z), \\
\label{align:MK_3}&\Prob\left[\Delta_W(z) = \delta_z\right] = \displaystyle\sum_{y} T_M(y, z) \Prob\left[\Delta_V(y) = \delta_y\right],
\end{align}
for all $y, z\in \mathcal{Y}$. Note that \eqref{align:MK_3} is equivalent to
\begin{equation}\label{eq:MK_4}
q_W(z) = \displaystyle\sum_{y} T_M(y, z) q_V(y),
\end{equation}
and from \eqref{align:MK_2}, we get
\begin{equation}\label{eq:MK_5}
V(y|0)-V(y|1) = \displaystyle\sum_{z} \widetilde{T}_M(y, z) \left(W(z|0)-W(z|1)\right),
\end{equation}
where 
\begin{equation}\label{eq:MK_6}
\widetilde{T}_M(y, z) = \displaystyle\frac{q_V(y)}{q_W(z)} T_M(y, z).
\end{equation}
Now, observe that via \eqref{eq:MK_6}, we have
\begin{equation*}
\displaystyle\sum_{y} \widetilde{T}_M(y, z) = 1.
\end{equation*}
Moreover, taking the denominator $q_W(z)$ in \eqref{eq:MK_6} to the other side, summing over $z$, and using \eqref{align:MK_1}, we get 
\begin{equation}\label{eq:MK_7}
V(y|0)+V(y|1) = \displaystyle\sum_{z} \widetilde{T}_M(y, z) \left(W(z|0)+W(z|1)\right).
\end{equation}
Combining \eqref{eq:MK_5} and \eqref{eq:MK_7} gives 
\begin{equation*}
V(y|x) = \displaystyle\sum_{z} \widetilde{T}_M(y, z) W(z|x),
\end{equation*}
for $x\in\{0, 1\}$. This proves that convex ordering implies stochastic degradation as 
$\widetilde{T}_M(y, z)$ is of the form of $P(y|z)$ given in \eqref{eq:sd}. This concludes the proof of the equivalence claim.

\subsection{Tools for Verifying the Symmetric Convex Ordering}
As the symmetric convex ordering between two channels can be described via the increasing convex ordering of their $\lvert\Delta_W\rvert$ parameters, we can borrow any tool from the literature used to verify the latter.    
In the next theorem, a `simple' criterion, known as the \textit{Karlin-Novikoff cut criterion} \cite{stanford1962generalized}, is given for two random variables to satisfy the increasing convex ordering\footnote[4]{We also note that a more general version of the cut criterion called 
Karlin-Novikoff-Stoyan-Taylor crossing conditions for stop-loss order can be found in \cite{Hurlimann2008}. }.
\begin{theorem}\cite[Theorem E]{szekli1995stochastic}\label{thm::cut_criterion}
Suppose that for $\Delta_1, \Delta_2$ with finite first moments $m_{\Delta_1} = \Expt[\Delta_1]$ and $m_{\Delta_2} = \Expt[\Delta_2]$, we have $m_{\Delta_1} \leq m_{\Delta_2}$ and
\begin{align}
\label{eq::cut_criterion_1}&F_{\Delta_1}(\delta) \leq F_{\Delta_2}(\delta), \quad \hbox{for } \delta \leq c,\\
\label{eq::cut_criterion_2}&F_{\Delta_1}(\delta) \geq F_{\Delta_2}(\delta), \quad \hbox{for } \delta > c,
\end{align}
for some $c\in \mathbf{R}$, then $\Delta_1\prec_{icx}\Delta_2$. 
\end{theorem}

The theorem provides a necessary and sufficient condition for the\textit{ stop-loss order} which is the name given to the increasing convex ordering in the actuarial science literature.

In the comparison process, the following idea will also be useful for checking our ordering.  
%We will use this idea as a tool to verify the new `symmetric convex/concave ordering' we introduce.
\begin{definition}\cite[Definition 1.3]{4461/THESES}\label{def:symmetrization}
For any B-DMC $W:\mathcal{X}\to\mathcal{Y}$, the \textit{symmetrized} B-DMC $W_s:\mathcal{X}\to\mathcal{Y}\times\mathcal{X}$ is defined as 
\begin{equation*}
W_s(y, z|x) = \frac{1}{2} W(y|x\oplus z).
\end{equation*}
\end{definition}

\subsection{Novelty of the Ordering by an Example}
We saw that any channel $V$ which satisfies the relation $\Delta_{V} \prec_{cx} \Delta_{W}$ with respect to any other channel $W$ is in fact stochastically degraded 
with respect to $W$. It is also clear by definition that the convex ordering between the channels implies the symmetric convex ordering introduced in Definition \ref{def:sym_cx_ordering}. 
So, we need to study the reverse implication to decide whether the symmetric convex ordering condition of
Theorem \ref{thm:icx} gives a strictly weaker condition than convex ordering (stochastic degradation). At this point, by recalling the equivalence stated in Proposition \ref{prop:cx_order_sym}, we notice that this is not the case 
for symmetric channels as the two orders $\Delta_{V} \prec_{cx} \Delta_{W}$ and
$\lvert\Delta_{V}\rvert \prec_{icx} \lvert\Delta_{W}\rvert$ are equivalent for symmetric channels. The purpose of this subsection is to show that no equivalence exists between the symmetric convex ordering and stochastic degradation 
if one of the two channels is asymmetric. If we can find a pair of B-DMCs that does not satisfy stochastic degradation, but satisfies the symmetric convex ordering, we will be done. Such a pair is illustrated in the next example.

\begin{figure}[!ht]
\centering
\scalebox{1}{\input{z-channel}}
\caption{W is a Z-Channel.}\label{fig:z-channel} 
\hspace{20mm}\\
\scalebox{1}{\input{b-s-channel}}
\caption{V is a Binary Symmetric Channel.}\label{fig:b-s-channel}
\end{figure}

\begin{example}\normalfont
Let $W$ be a Z-channel with crossover probability $r\in[0, 1]$ and $V$ be a binary symmetric channel with crossover probability $p\in[0, 0.5]$. The channels are shown in Figure \ref{fig:z-channel} and Figure \ref{fig:b-s-channel}, respectively.
In this example, we will answer the following three questions:
\begin{enumerate}
 \item[($q1$)] Suppose $V$ is a stochastically degraded version of $W$. What is the best possible binary symmetric channel (with the smallest $p$) which satisfies this condition?
 \item[($q2$)] Suppose instead that the channels satisfy the symmetric convex ordering $\lvert\Delta_{V}\rvert \prec_{icx} \lvert\Delta_{W}\rvert$. What is the best possible binary symmetric channel which satisfies this condition?
 \item[($q3$)] Suppose we first symmetrize $W$ according to Definition \ref{def:symmetrization} to construct $W_s$. Suppose now $V$ is a stochastically degraded version of $W_s$. What is the best possible binary symmetric channel which satisfies this condition? 
\end{enumerate}
Then, we will compare the three binary symmetric channels to decide which ordering results in a better channel with a smaller crossover probability $p$, and thus leads to a polar code with a larger information subset over the Z-channel. Note that the information sets of the polar codes designed for each of the binary symmetric channels are all subsets of the information set of the capacity achieving polar code designed for the Z-channel. Thus, the polar code designed for the binary symmetric channel with a smaller crossover probability will achieve a larger rate over the Z-channel. Here are the answers.

($a1$) \textit{Stochastic degradation: } 
 Let us derive the range of possible values of $p$ in terms of $r$ under this assumption. 
For this purpose, we define the asymmetric binary channel $P$ degrading $W$ to $V$ by 
\begin{equation}\label{eq:deg_cond}
 V(y|x) = \displaystyle\sum_{z\in\{0, 1\}} W(z|x) P(y|z).
\end{equation}
First we note that $P(0|0) = 1-p$ and $P(0|1) = p$ are the only possibilities. Let $P(0|1) = \alpha$. Then, using \eqref{eq:deg_cond}, we get
\begin{equation*}
V(0|1) = p = (1-r)\alpha + r (1-p),
\end{equation*}
which implies 
\begin{equation}\label{eq:ex_1}
p = \displaystyle\frac{r+(1-r)\alpha}{1+r}.
\end{equation}
Noting that the right hand side of \eqref{eq:ex_1} is increasing in $\alpha\in[0,1]$, we conclude that
\begin{equation*}
 \displaystyle\frac{r}{1+r}\leq p \leq \displaystyle\frac{1}{1+r}
\end{equation*}
whenever we impose stochastic degradation on the channels. Picking the binary symmetric channel having the smallest crossover probability $p = r/(1+r)$ answers the first question. 

($a2$) \textit{$\lvert\Delta_{V}\rvert \prec_{icx} \lvert\Delta_{W}\rvert$: }
Now, we will derive the range of possible values of $p$ in terms of $r$ under this assumption by using the cut-criterion given in Theorem \ref{thm::cut_criterion}. 
We start by computing the values of $E[\lvert\Delta_{V}\rvert]$ and $E[\lvert\Delta_{W}\rvert]$ in terms of the channel parameters. For the binary symmetric channel, we have $E[\lvert\Delta_{V}\rvert] = 1-2p$. For the Z-channel, we have 
\begin{equation}
|\Delta_W(y)| = \begin{cases}
\displaystyle\frac{1-r}{1+r},& \hbox{if } y = 0\\
1,& \hbox{if } y = 1
\end{cases},
\end{equation}
$q_W(0) = (1+r)/2$, and $q_W(1) = (1-r)/2$. So, we compute $E[\lvert\Delta_{W}\rvert] = 1-r$. Note that any B-DMC together with any binary symmetric channel with crossover probability $p$ will always satisfy 
the conditions \eqref{eq::cut_criterion_1} and \eqref{eq::cut_criterion_2} of Theorem \ref{thm::cut_criterion} for $c = \lvert 1-2p\rvert$ and $F_{\Delta_1}$ corresponding to the cumulative distribution of the binary symmetric channel. As a result, we can see by the theorem's statement that the condition $E[\lvert\Delta_{V}\rvert] \leq E[\lvert\Delta_{W}\rvert]$ is a necessary condition in our example for $\lvert\Delta_{V}\rvert\prec_{icx}\lvert\Delta_{W}\rvert$ to hold. 
This in turn implies that $p \geq r/2$. Hence, 
the best possible binary symmetric channel in this case has crossover probability $p = r/2$. This answers the second question. 

($a3$) \textit{Channel symmetrization: } 
We first note a more general result: a given B-DMC $W'$ and its symmetrized version $W_s'$ always satisfy $|\Delta_{W_s'}(y, z)| = |\Delta_{W'}(y)|$ with $|\Delta_{W_s'}(y, z)|$ distributed as $0.5q_{W'}(y)$, for $z=\{0, 1\}$. Therefore, for any function $f(\delta)$ defined for $\delta\in[0,1]$, we have
\begin{equation*}
\Expt\left[f(|\Delta_{W'}|)\right] = \Expt\left[f(|\Delta_{W_s'}|)\right].
\end{equation*} 
We conclude that for any two B-DMCs $W'$ and $V'$: $|\Delta_{V'}| \prec_{icx} |\Delta_{W'}|$ if and only if $|\Delta_{V_s'}| \prec_{icx} |\Delta_{W_s'}|$. Moreover, as the channels in this last condition are symmetric, we know the condition holds if and only if $\Delta_{V_s'} \prec_{cx} \Delta_{W_s'}$, i.e., the symmetrized versions of the channels are ordered by stochastic degradation. So, we have the same answer as in the previous case: the best possible binary symmetric channel in this case has also crossover probability $p = r/2$.

Let us compare the results. Noting that $r/2 \leq r/(1+r)$ holds for any $r\in[0, 1]$, and with equality if and only if $r= \{0,1\}$, we conclude that, for $r\in(0, 1)$, the binary symmetric channel with smallest crossover probability is found by the symmetric convex ordering and this binary symmetric channel is not stochastically degraded with respect to the Z-channel. For instance, when $r=0.5$, the crossover probabilities of the best binary symmetric channel we found in the second case is $0.25$ compared to $1/3$ in the first one. Finally, we also showed that one can verify the symmetric convex ordering by first symmetrizing the asymmetric channels and then checking for stochastic degradation. The example proves that for general B-DMCs the symmetric convex ordering is strictly weaker than stochastic degradation.
\end{example} 

\subsection{Squeezing the Information Sets Between Binary Erausre Channels}
Finally, we discuss two other orderings related to binary erasure channels. Recall that we provided the definitions of the Bhattacharyya parameter $Z(W)$ and the symmetric capacity $I(W)$ of a B-DMC in \eqref{eq:Bhattacharyya} and \eqref{eq:sym_cap}, respectively, and we defined $T(W) = \Expt[\lvert\Delta_W\rvert]$. First we note the following property of the binary erasure channel.

\begin{proposition}\label{prop::BEC_max_min_I_Z}
Amongst the set of symmetric B-DMCs with a given fixed value of the channels' variational distance $T$ between their own transition probabilities, the binary erasure channel $U$ of erasure probability $1-T(U)$ 
maximizes the symmetric capacity and minimizes the Bhattacharyya parameter.
\end{proposition}
\begin{proof}
The proof follows by noting $T(U) = 1 - Z(U)$ for the binary erasure channel and using the following upper bounds to the uncoded error 
probability $(1-T(W))/2 \leq (1-I(W))/2$ and $(1-T(W))/2 \leq Z(W)/2$.
\end{proof}

For a channel $W$, we define\footnote[5]{Note that this is not the Bhattacharyya parameter $Z(W)$.} $Z_W = \lvert\Delta_W\rvert$. Suppose a binary erasure channel $BEC$ with erasure probability $\epsilon\in[0,1]$ and a B-DMC $W$ satisfy $\Expt[Z_W] \leq E[Z_{BEC}]$. 
Note that $Z_{BEC}$ is $\{0, 1\}$ valued and satisfies $P(Z_{BEC} = 0) = \epsilon$. As a result, the random variable $Z_{BEC}$ and 
any arbitrary random variable $Z$ will satisfy the conditions \eqref{eq::cut_criterion_1} and \eqref{eq::cut_criterion_2} of Theorem \ref{thm::cut_criterion} when $\Delta_2$ is taken as the random variable $Z_{BEC}$ and $F_{\Delta_2}$ stands for its cumulative distribution. 
As a result, the assumption $T(W) = \Expt[Z_W] \leq E[Z_{BEC}] = T(W_{BEC})$ implies $Z_W \prec_{icx} Z_{BEC}$. By Theorem \ref{thm:icx}, we know that this ordering is preserved under the polar transform.

Another instance of the increasing convex ordering slightly different than Theorem \ref{thm:icx} happens when $BEC$ and $W$ are such that the Bhattacharyya parameters of the channels satisfy $Z(W) \leq Z(BEC)$. 
Let us define the random variable $B_W =  \sqrt{1-Z_W^2}$. Then, $Z(W) = \Expt[B_W]$. 
Hence, the channels satisfy $\Expt[B_W] \leq \Expt[B_{BEC}]$. Letting this time the random variable $\Delta_2$ in Theorem \ref{thm::cut_criterion} stand for the random variable $B_{BEC}$ and $F_{\Delta_2}$ for its cumulative distribution, we see that $\Expt[B_W] \leq \Expt[B_{BEC}]$ implies $B_W \prec_{icx} B_{BEC}$. Finally, it is well known from \cite[Proposition 6]{1669570} that this ordering is also preserved under the polar transform.  

Using these two binary erasure channel orderings, the following theorem shows that the information set of a given symmetric B-DMC can be squeezed between the information sets of two binary erasure channels.
\begin{theorem}\label{thm:info_set_order}
For any given symmetric B-DMC $W$ with parameter values $T(W)$ and $Z(W)$, define the binary erasure channel $U$ such that $T(U) = T(W)$ and the binary erasure channel $V$ such that $Z(V) = Z(W)$. Then, we have 
$Z(U^{s^n})\leq Z(W^{s^n}) \leq Z(V^{s^n})$ for any $s^n\in\{+, -\}^n$ with $n = 0, 1, \ldots$ Furthermore, this implies the following ordering of the information sets:
\begin{equation*}
\mathcal{A}_N^{f_s, \epsilon}(V)\subseteq\mathcal{A}_N^{f_s, \epsilon}(W)\subseteq\mathcal{A}_N^{f_s, \epsilon}(U), \quad \forall\epsilon\in[0, 1],
\end{equation*}
for $N=2^n$ and the function $f_s(\delta) = 1 - \sqrt{1 - \delta^2}$.
\end{theorem}
\begin{proof}
It is already known that the binary erasure channel $V$ provides universally good indices: $\mathcal{A}_V \subseteq \mathcal{A}_W$\cite{1669570}. To prove the other claim, we first note that the following extremality results hold by \cite[Proposition 4]{6621586}:
\begin{align*}
 T(W^-) &= T(W)^2, \\
 T(W^+) &\in \left[T(W), 2T(W) - T(W)^2\right].
\end{align*} 
The proof that the binary erasure channel $U$ provides universally bad indices follows by Proposition \ref{prop::BEC_max_min_I_Z} and the above extremality result, upon noticing that being a binary erasure channel is preserved under the polarization transformations with $T(U^+) = 2T(U) - T(U)^2$.
\end{proof}

\section{Polarization Property}\label{sec:pol_property}
The following lemma proves that the polarization property of the polar transform holds for all the channel parameters $T_{f_s}(W)$, where $f_s\in\mathcal{F}_{s, cx}$.
\begin{lemma}\label{lem:pol_property} For any two B-DMCs $W_1$ and $W_2$,we have
\begin{align*}
T_{f_s}(W_{1, 2}^-) \leq T_{f_s}(W_1) \leq T_{f_s}(W_{1, 2}^+),  \\
T_{f_s}(W_{1, 2}^-) \leq T_{f_s}(W_2) \leq T_{f_s}(W_{1, 2}^+),
\end{align*}
for any $f_s\in\mathcal{F}_{s, cx}$.
\end{lemma}
\begin{proof}
The idea behind the proof of this lemma is exactly the same idea used in \cite[Proof of Lemma 3]{6731577}. First, note that the channels $W_{1, 2}^\pm$ and $W_{2, 1}^\pm$ have the same $T_{f_s}$ values. Thus, it would be sufficient to show the first set of inequalities. 

As for any realizations $\delta_1$ and $\delta_2$ of the random variables $\Delta_{W_1}(y_1)$ and $\Delta_{W_2}(y_2)$, respectively, 
$\lvert \delta_1\delta_2\rvert\leq\lvert \delta_1\rvert$ holds, we have 
\begin{equation*}
f_s (\delta_1\delta_2) \leq f_s(\delta_1),
\end{equation*}
for any $f_s\in\mathcal{F}_{s, cx}$. Taking expectations of both sides, we get $T_{f_s}(W_{1, 2}^-) \leq T_{f_s}(W_1)$.

On the other side, we have
\begin{equation*}
\displaystyle\frac{1 + \delta_1\delta_2}{2} f_s\left(\displaystyle\frac{\delta_1 + \delta_2}{1 + \delta_1\delta_2}\right) 
+ \displaystyle\frac{1 - \delta_1\delta_2}{2} f_s\left(\displaystyle\frac{\delta_1 - \delta_2}{1 - \delta_1\delta_2}\right)
\geq f_s\left(\displaystyle\frac{\delta_1 + \delta_2}{2} + \displaystyle\frac{\delta_1 - \delta_2}{2} \right) = f_s(\delta_1),
\end{equation*}
by Jensen's inequality. Taking expectations, $T_{f_s}(W_1) \leq T_{f_s}(W_{1, 2}^+)$ follows.
\end{proof}
Using Definition \ref{def:sym_cx_ordering}, the following corollary follows from the lemma.
\begin{corollary}
The channels $W_{1, 2}^-$, $W_1$, $W_{1, 2}^+$ satisfy the symmetric convex ordering:
\begin{equation*}
\big\lvert\Delta_{W_{1, 2}^-}\big\rvert \prec_{icx} \lvert\Delta_{W_1}\rvert \prec_{icx} \big\lvert\Delta_{W_{1, 2}^+}\big\rvert. 
\end{equation*}
The same result holds for the channel $W_2$.
\end{corollary}

\section{Applications to Polar Coding over Non-Stationary B-DMCs}\label{sec:applications}
The original theory of polar coding is extended to non-stationary B-DMCs in \cite{6874843}. It is shown there that the recursive application of the generalized polar transform (which allow to combine and split arbitrary independent channels) polarizes non-stationary memoryless channels in the same way the polar transform polarizes stationary ones. In this section, we discuss two applications of the order preserving property of the generalized polar transform shown in Theorem \ref{thm:icx}.

\subsection{Efficient Construction of the Information Sets of Polar Codes}
In the beginning of the paper, we acknowledged the difficulty in computing efficiently the exact transition probabilities of the synthetic channels when these have very large output alphabets. Here, we make a quick look into how, despite this underlying difficulty, the information sets of polar codes can still be efficiently constructed. The idea of the approximation algorithm used in \cite{6557004} for the stationary setting can be summarized as follows: 
Once the output alphabets of the the synthetic channels become too large, they are replaced by channels $(i)$ which are `close' to the original channels, $(ii)$ which have permissible output alphabet sizes, and $(iii)$ whose children synthesized by the sequence of polar transformations still remain `close' to their exact versions. Thus, the key point is to use an approximation algorithm inducing an ordering which is preserved by the polar transform. In \cite{6557004}, stochastic degradation is used for that purpose, and it is shown that the algorithm performs well ---a further analysis of the algorithm carried out in \cite{6033724} bounds the maximum approximation loss of the algorithm and shows that the algorithm works with almost linear complexity in the block-length. 

As the symmetric convex ordering is a (weaker) partial order also preserved by the polar transform, it can be used as an alternative approximation method for the asymmetric synthetic channels. Although we have not implemented such an algorithm to evaluate its performance, we claim that similar guarantees can be obtained given the fact that both convex ordering (stochastic degradation) and symmetric convex ordering are induced via the fusion (merging) of the outputs. By Theorem \ref{thm:info_set_order}, we can easily see how the exact and approximate computations can be abandoned once the gap between the information sets of the two specific binary erasure channels defined in the theorem's statement is sufficiently small. In that case, the algorithm proceeds by using the binary erasure channel recursion for some channel parameters such as the Bhattacharyya distance, and eventually terminate.  

More importantly, we claim that the results can be extended to non-stationary memoryless channels. As it is shown in \cite{6874843} that a construction combining non-identical channels with the polar transform does still make sense, we believe that the idea of the algorithm proposed in \cite{6557004} should remain useful for approximating the transition probabilities of the synthetic channels in the non-stationary setting. In particular, we claim that in the non-stationary setting the symmetric convex ordering can be applied in order to efficiently approximate and reduce the output alphabet sizes of both the symmetric and asymmetric channels synthesized by the sequence of generalized polar transformations. 

\subsection{Universal Polar Coding with Channel Knowledge at the Decoder}
In the introduction, we referred to an important problem related to the design of polar codes, namely the size of the intersection $\mathcal{A}_{N}(W)\cap \mathcal{A}_{N}(V)$ for two given B-DMCs $W$ and $V$. This information would be highly useful to a code designer who wants to use the polar code designed for one of the channels over the other one. Here, we are only interested in using the original polar code design of Ar{\i}kan \cite{1669570} and leave out any derivative design (in the stationary setting) which are out of this scope. In the next corollary, we show that the symmetric convex ordering induces the subset ordering for the information sets of polar codes over the non-stationary memoryless B-DMCs it orders, and thus the smallest of the information sets can be used for reliable communication over all of the ordered channels\footnote[6]{Note that this will result in a code with a rate smaller than the capacities of all of the ordered channels, except the design channel.}.

Before we start the discussion, we need to introduce some notations from \cite{6874843}. 
Suppose $W_t$ is the channel law at time instant $t\in\N$. For a given block-length $N = 2^n$ with $n = 0, 1, \ldots$, each stage of Arıkan's polar construction applying the generalized polar transform will successively transform this collection of channels into a collection $\{W_{k,t}\colon t\in\N\}$ of channels, where $k = 0, \ldots, n$ indicates the corresponding stage of the recursion.

\begin{corollary}\label{cor:info_set_order_non_stationary}
Let $\mathcal{W}$ be a set of B-DMCs and $V$ be a B-DMC such that
$$
\lvert\Delta_{V}\rvert \prec_{icx} \lvert\Delta_{W}\rvert,
$$
for all $W\in\mathcal{W}$. Then, the polar code designed for the channel $V$ is universal for the class $\mathcal{W}$ in the sense that if $W_{0,t}\in\mathcal{W}$, for any $t\in\N$, the following subset orderings hold: 
\begin{equation}\label{eq:non-stat-subset}
\mathcal{A}_N^{f_s, \epsilon}(V) \subseteq \mathcal{A}_N^{f_s, \epsilon}\left(\{W_{n,t}\colon t\in\N\}\right),
\end{equation}
for any $N = 2^n$ with $n = 0, 1, \ldots$, any $f_s\in\mathcal{F}_{s, cx}$, and any $\epsilon\in(0, 1)$.
\end{corollary} 
\begin{proof}
The result follows as a corollary to Theorem \ref{thm:icx}. For notational consistency, we denote by $\{V_{n,t}\colon t\in\N\}$ the set of synthetic channels obtained from the $n$-fold application of the polar transform to copies of the channel $V$, i.e, we have $V_{0, t} = V$, for any $t\in\N$. By the preservation property shown in Theorem \ref{thm:icx} and the recursive construction procedure, we conclude that  
$$
\big\lvert\Delta_{V_{n, t}}\big\rvert \prec_{icx} \big\lvert\Delta_{W_{n,t}}\big\rvert,
$$
hold for all $n=0, 1, \ldots$ and any $t\in\N$. From this relation, the claim in \eqref{eq:non-stat-subset} follows. 
\end{proof}
Assuming that the decoder knows the sequence of realizations of the non-stationary memoryless channel, the corollary reveals that the universality arising from the symmetric convex ordering, and hence from stochastic degradation, extends form the stationary setting to the non-stationary one.

\section{Final Remarks}\label{sec:final-rem}
This paper proposed the symmetric convex ordering as a novel partial ordering for communication channels. The study revealed that this ordering is a strictly weaker partial ordering than stochastic degradation and leads to the subset ordering of the information sets of polar codes. The subset ordering is a consequence of Theorem \ref{thm:icx} which shows that the polar transform preserves symmetric convex orderings. This final section closes the paper by highlighting the novelty of our results in the light of the previous literature.

It was brought to the author's attention that in the LDPC coding literature a well-known result for symmetric channels states that stochastic degradation is equivalent to the increasing convex ordering of $|D|$-densities, see for instance \cite[Theorem 4.76]{Richardson:2008:MCT:1795974}. However, the term (increasing) convex ordering seems not to have been adopted by the researchers in the field, even though the theory of stochastic orders likely pre-dates sources such as \cite{Richardson:2008:MCT:1795974}. Therefore, one distinctive quality of this paper is the description of connections with the theory of stochastic orders. 

In addition, up to our knowledge, the mentioned equivalence property has not been investigated before for asymmetric channels. In that respect, this work contributes to the literature by showing that such an equivalence does not hold if one of the channels is asymmetric. The readers familiar with the coding and information theory literature might argue that the justification for not asking about what happens to the equivalence in the case of asymmetric channels follows from the channel symmetrization argument\footnote[7]{The origins of this argument are not clear to this author, but the argument has been used by multiple researchers in the field.}. Let us explain what is missing in this approach. As far as we know, channel symmetrization has been used in these fields as an argument to extend the results derived for symmetric channels\footnote[8]{i.e., those derivations facilitated by the use of the symmetry property of a symmetric  channel.} to asymmetric ones. The extension becomes straightforward after realizing that the performance measures of asymmetric channels and their symmetrized versions have the same value when evaluated under the uniform input distribution. For instance, the idea has been used in \cite[Lemma 1.4]{4461/THESES} in the context of source polarization and polar codes. The important point to notice is the following: as opposed to this work, in these contexts channel symmetrization was not explicitly viewed as a tool for ordering the channels. One can see this more concretely by carefully looking at Example 1, so let use explore the idea of that example. Suppose that we have an asymmetric channel $W$ and we would like to find the set of channels which are degraded with respect to $W$. Call this set $S_1$ . Then, suppose
we symmetrize the channel to obtain $W_s$, and we find the set of channels which are degraded with respect to $W_s$. Call this set $S_2$. Now, we ask the following question: Is $S_1 = S_2$ , and why or why not? This question we pose is critical in understanding how this work distinguishes itself from the previous works. 
As we know now, this manuscript shows that in general $S_1 \subset S_2$, i.e., stochastically degrading the symmetrized version $W_s$ will result in a larger set of channels, and explains the reason: symmetrizing the channel and using stochastic degradation is equivalent to degrading the channel in the sense of the symmetric convex ordering, and this latter ordering is a strictly weaker partial ordering than stochastic degradation. Therefore, another merit of this paper is in its application of channel symmetrization.

Another point worth commenting is related to the definition of the family of more general information measures we denoted by $T_{f_s}(W)$, for $f_s\in\mathcal{F}_{cx,s}$. The idea of studying a more general family of information measures is not a new one and has been a subject of interest to many information theorists. One general family of information measures is the family of f-divergences: the notion of divergence between two probability distributions $p$ and $q$ was generalized to f-divergences by the authors of \cite{Csis63} and \cite{ali_66_general} as follows:
\begin{equation}
D_f(p||q) = \displaystyle\sum_{z\in\mathcal{Z}} p(z) f\left(\displaystyle\frac{q(z)}{p(z)}\right),
\end{equation}
where $f(.)$ is a convex function. For instance, a well known f-divergence is the variational distance $T(W)$ between $p \leftarrow W(y|0)$ and $q \leftarrow W(y|1)$. Therefore, one could ask regarding our definition whether the measures $T_{f_s}(W)$ are all f-divergences. For simplicity, let us look at the case of a symmetric channel $W$. In this case, the expression simplifies to
\begin{multline}
T_{f_s}(W)=\Expt[f_s(\Delta(W))] = \displaystyle\sum_{y\in\mathcal{Y}} \displaystyle\frac{W(y|0) + W(y|1)}{2} f_s\left(\frac{W(y\mid0)-W(y\mid1)}{W(y\mid0)+W(y\mid1)}\right)\\
=  \displaystyle\sum_{y\in\mathcal{Y}} W(y|0) f_s\left(h\left(\frac{W(y\mid1)}{W(y\mid0)}\right)\right),
\end{multline}
by using the symmetry property of the channel and the symmetry of the function $f_s$ and defining  
\begin{equation}
h(t) = \displaystyle\frac{1-t}{1+t},
\end{equation}
for $t\in [0, \infty)$. By assumption, the function $f_s(.)$ is a convex and symmetric function in $[-1, 1]$ (decreasing in $[-1, 0]$ and increasing in $[0, 1]$), and one can verify that the function $h$ is a convex decreasing function in $[0, \infty)$. Now, one can easily check that under these assumptions the second derivative of the composite function $f_s(h(t))$ is not necessarily non-negative, i.e., the composite function is not necessarily a convex function in $t\in [0, \infty)$. As a result, we conclude that the measures we focused in this paper ---$T_{f_s}(W)$ with the functions $f_s$ belonging to the family of convex and symmetric functions $\mathcal{F}_{cx,s}$--- are not necessarily f-divergences.
 
Finally, we note that other partial orderings for communication channels that are strictly weaker than stochastic degradation have been proposed  in the literature. K{\"o}rner and Marton defined the less noisy and more capable channels in \cite{less_noisy}, and an application of these orderings in the context of broadcast channels can be found in \cite{1056029}. We note the work in \cite{6875075} which study the subset ordering problem for the information sets of polar codes over stationary memoryless channels and show that the less noisy ordering also implies the subset ordering for the information sets of polar codes.

\section*{Acknowledgment}
This work was supported by the Swiss National Science Foundation under grant number 200021-125347/1. The author would like to thank the anonymous Reviewers for their valuable comments. 

\section*{Appendix}
In this appendix, we state and prove Lemma \ref{lem:f_plus_convex}

\begin{lemma}\label{lem:f_plus_convex}
 Let $f_s(\delta)$ be a convex and symmetric function in $\delta\in[-1, 1]$. Then, the function defined in \eqref{eq:f_plus} is also a convex and symmetric function.
\end{lemma}
\begin{proof}
For simplicity, we first define 
\begin{align*}
 f_1(\delta_1, \delta_2) &= (1 + \delta_1 \delta_2) f_s\left( \frac{\delta_1 + \delta_2}{1 + \delta_1 \delta_2}\right),  \\
 f_2(\delta_1, \delta_2) &= (1 - \delta_1 \delta_2) f_s\left( \frac{\delta_1 - \delta_2}{1 - \delta_1 \delta_2}\right),
\end{align*}
$\delta_1, \delta_2\in[-1, 1]$.
Hence, \eqref{eq:f_plus} equals to
\begin{equation*}
 f^+(\delta_1, \delta_2) = \frac{1}{2} f_1(\delta_1, \delta_2) + \frac{1}{2} f_2(\delta_1, \delta_2).
\end{equation*}
As $f^+(\delta_1, \delta_2) = f^+(\delta_2, \delta_1)$, it is sufficient to prove the lemma for one of the variables. One can easily prove that the function is symmetric in $\delta_1\in[-1, 1]$, i.e, $f^+(\delta_1, \delta_2) = f^+(-\delta_1, \delta_2)$ by using the symmetry of the function $f_s(\delta)$ in $\delta\in[-1, 1]$.

We will prove the rest of the lemma for smooth functions $f_s$. As such functions are dense, this is without loss of generality. Let $f_s''$ denote the second derivative of $f_s(\delta)$ with respect to the variable $\delta$. Then, we get
\begin{equation*}
 \frac{\partial}{\partial \delta_1} f_1(\delta_1, \delta_2) = \delta_2 f_s\left( \frac{\delta_1 + \delta_2}{1 + \delta_1 \delta_2}\right) + \frac{1 - {\delta_2}^{2}}{1 + \delta_1 \delta_2} f_s'\left( \frac{\delta_1 + \delta_2}{1 + \delta_1 \delta_2}\right),  
\end{equation*}
\begin{align*}
 \frac{\partial^2}{\partial {\delta_1}^{2}} f_1(\delta_1, \delta_2) &= \delta_2 f_s'\left( \frac{\delta_1 + \delta_2}{1 + \delta_1 \delta_2}\right) \frac{1 - {\delta_2}^{2}}{(1 + \delta_1 \delta_2)^2} \\
 &- \delta_2 \frac{1 - {\delta_2}^{2}}{(1 + \delta_1 \delta_2)^2}f_s'\left( \frac{\delta_1 + \delta_2}{1 + \delta_1 \delta_2}\right) \\
 &+ \frac{1 - {\delta_2}^{2}}{1 + \delta_1 \delta_2} f_s''\left( \frac{\delta_1 + \delta_2}{1 + \delta_1 \delta_2}\right) \frac{1 - {\delta_2}^{2}}{(1 + \delta_1 \delta_2)^2} \\
 &= \frac{(1 - {\delta_2}^{2})^2}{(1 + \delta_1 \delta_2)^3} f_s''\left( \frac{\delta_1 + \delta_2}{1 + \delta_1 \delta_2}\right).   
\end{align*}
Similarly, we get 
\begin{equation*}
 \frac{\partial}{\partial \delta_1} f_2(\delta_1, \delta_2) 
 = - \delta_2 f_s( \frac{\delta_1 - \delta_2}{1 - \delta_1 \delta_2}) + \frac{1 - {\delta_2}^{2}}{1 - \delta_1 \delta_2} f_s'( \frac{\delta_1 - \delta_2}{1 - \delta_1 \delta_2}),   
\end{equation*}
and
\begin{equation*}
 \frac{\partial^2}{\partial {\delta_1}^{2}} f_2(\delta_1, \delta_2) =  \frac{(1 - {\delta_2}^{2})^2}{(1 - \delta_1 \delta_2)^3} f_s''( \frac{\delta_1 - \delta_2}{1 - \delta_1 \delta_2}). 
\end{equation*}
Summing these we obtain 
\begin{equation*}
 \frac{\partial^2}{\partial {\delta_1}^{2}} f^+(\delta_1, \delta_2) = \frac{1}{2}\frac{(1 - {\delta_2}^{2})^2}{(1 + \delta_1 \delta_2)^3} f_s''\left( \frac{\delta_1 + \delta_2}{1 + \delta_1 \delta_2}\right) 
 + \frac{1}{2}\frac{(1 - {\delta_2}^{2})^2}{(1 - \delta_1 \delta_2)^3} f_s''\left( \frac{\delta_1 - \delta_2}{1 - \delta_1 \delta_2}\right) \geq 0,
 %\frac{\partial^2}{\partial {\delta_2}^{2}} f^+(\delta_1, \delta_2) &= \frac{1}{2}\frac{(1 - {\delta_1}^{2})^2}{(1 + \delta_1 \delta_2)^3}\hspace{2mm} f''( \frac{\delta_1 + \delta_2}{1 + \delta_1 \delta_2}) + \frac{1}{2}\frac{(1 - {\delta_1}^{2})^2}{(1 - \delta_1 \delta_2)^3}\hspace{2mm} f''( \frac{\delta_1 - \delta_2}{1 - \delta_1 \delta_2}) \geq 0  
\end{equation*}
where the sign of $f^+(\delta_1, \delta_2)$ can be deduced from the convexity of the function $f_s(\delta)$ in $\delta\in[-1, 1]$. This proves that $f^+(\delta_1, \delta_2)$ is convex in $\footnote[5]{Notwe that this is not the Bhattacharyya parameter $Z(W)$!}\delta_1\in[-1, 1]$ and completes the proof. 
\end{proof}

\bibliographystyle{IEEEtran}
\bibliography{ref}

\end{document}

%% file: z-channel.tex
\setlength{\unitlength}{1bp}%
\begin{picture}(140,100)
\small
\put(0,0){\includegraphics[scale=1.0]{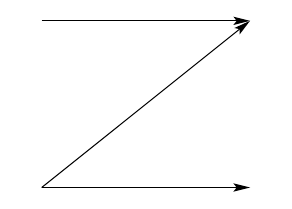}}
\put(10,10){\makebox(0,0){$1$}}
\put(10,90){\makebox(0,0){$0$}}
\put(130,10){\makebox(0,0){$1$}}
\put(130,90){\makebox(0,0){$0$}}
\put(70,2){\makebox(0,0){$1-r$}}
\put(45,40){\makebox(0,0){$r$}}
\put(70,98){\makebox(0,0){$1$}}
\end{picture}

%% file: b-s-channel.tex
\setlength{\unitlength}{1bp}%
\begin{picture}(140,100)
\small
\put(0,0){\includegraphics[scale=1.0]{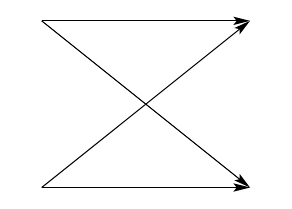}}
\put(10,10){\makebox(0,0){$1$}}
\put(10,90){\makebox(0,0){$0$}}
\put(130,10){\makebox(0,0){$1$}}
\put(130,90){\makebox(0,0){$0$}}
\put(70,2){\makebox(0,0){$1-p$}}
\put(45,40){\makebox(0,0){$p$}}
\put(45,60){\makebox(0,0){$p$}}
\put(70,98){\makebox(0,0){$1-p$}}
\end{picture}